\documentclass[tightenlines,superscriptaddress,floatfixolumn,preprint,nofootinbib]{revtex4-1} 
\usepackage{graphicx}
\usepackage{amsmath}
\usepackage{amsfonts,amsbsy}
\usepackage{amssymb}
\usepackage{xfrac}
\usepackage{hyperref}

\begin{document}
\title{Probing proton fluctuations with asymmetric rapidity correlations}
\author{Adam Bzdak}
\affiliation{AGH University of Science and Technology,
Faculty of Physics and Applied Computer Science, 30-059 Krak\' ow, Poland}
\email{bzdak@fis.agh.edu.pl}

\author{Kevin Dusling}
\affiliation{American Physical Society, 1 Research Road, Ridge, NY 11961, USA}
\email{kdusling@mailaps.org}

\begin{abstract}
Intrinsic fluctuations of the proton saturation momentum generate asymmetric
rapidity distributions on an event-by-event basis.  We argue that the
asymmetric component, $\left<a_1^2\right>$, of the orthogonal polynomial
decomposition of the two-particle rapidity correlation function is a sensitive
probe to this distribution of fluctuations.  We present a simple model
connecting the experimentally measured $\left<a_1^2\right>$ to the variance,
$\sigma$, of the distribution of the logarithm of the proton saturation scale.
We find that $\sigma\approx 0.5-1$ describes the asymmetric component of the
rapidity correlations recently measured by the ATLAS collaboration.
\end{abstract}

\maketitle
\def\Sp{S_\perp}
\newcommand{\Q}[1]{Q_{o,#1}^{2}}
\newcommand{\rh}[1]{{\rho_{#1}}}
\def\rhcm{{\bar{\rho}}}
\def\Nch{N_{\rm ch}}
\newcommand{\QmeanSq}[1]{\bar{Q}_{#1}^{\:2}}

\section{Introduction}

There has been considerable interest in the study of collective phenomena in
small colliding systems since the initial discovery of long-range azimuthally
collimated di-hadron correlations in high multiplicity proton-proton collisions
at the LHC \cite{Khachatryan:2010gv}. While the systematics of the measured
\emph{ridge-like} correlations have been confronted by both hydrodynamic and
Color Glass Condensate inspired models (see \cite{Dusling:2015gta} for a recent
review) the origin and nature of the high multiplicity events are far from
being understood. The strength of the near-side azimuthal correlation is
observed to grow monotonically with event activity up to the highest measured
multiplicities having about ten times the mean number of charged tracks which can 
only be accounted for by rare fluctuations of the proton wave function.

A recent work has shown that a combination of impact parameter and color charge
fluctuations are insufficient in describing the broad width of multiplicity
distribution in p+p collisions \cite{Schenke:2013dpa}. In order to accommodate
the data additional event-by-event fluctuations of the saturation scale in the
proton must be included. While this goes beyond the conventional Color Glass
Condensate framework, the existence of such fluctuations has been known for
almost a decade \cite{Munier:2003vc,Munier:2003sj,Iancu:2004es,Marquet:2006xm} as recently
emphasized in \cite{McLerran:2015lta} where it was shown that saturation scale
fluctuations of the proton are required to understand the centrality
dependence of the charged particle rapidity distribution in p+Pb collisions. By
including event-by-event fluctuations of the initial proton saturation momenta
within classical Yang-Mills simulations the authors of \cite{McLerran:2015qxa}
were able to describe the broad tail of the multiplicity distribution lending
credence to the importance of rare fluctuations in generating dense nuclear
configurations. 
 
The goal of this work is to demonstrate that the two-particle correlation
function in rapidity, $C_2(y_1,y_2)$, can serve as a sensitive probe of the
distribution of saturation scale fluctuations of the proton. It provides
constraints on the variance, $\sigma$, of the logarithm of the saturation scale
about its mean ({\em i.e.} minimum bias) position due to fluctuations occurring
in the tail of the dipole scattering amplitude.

\begin{figure}[h!]
\includegraphics[scale=0.8]{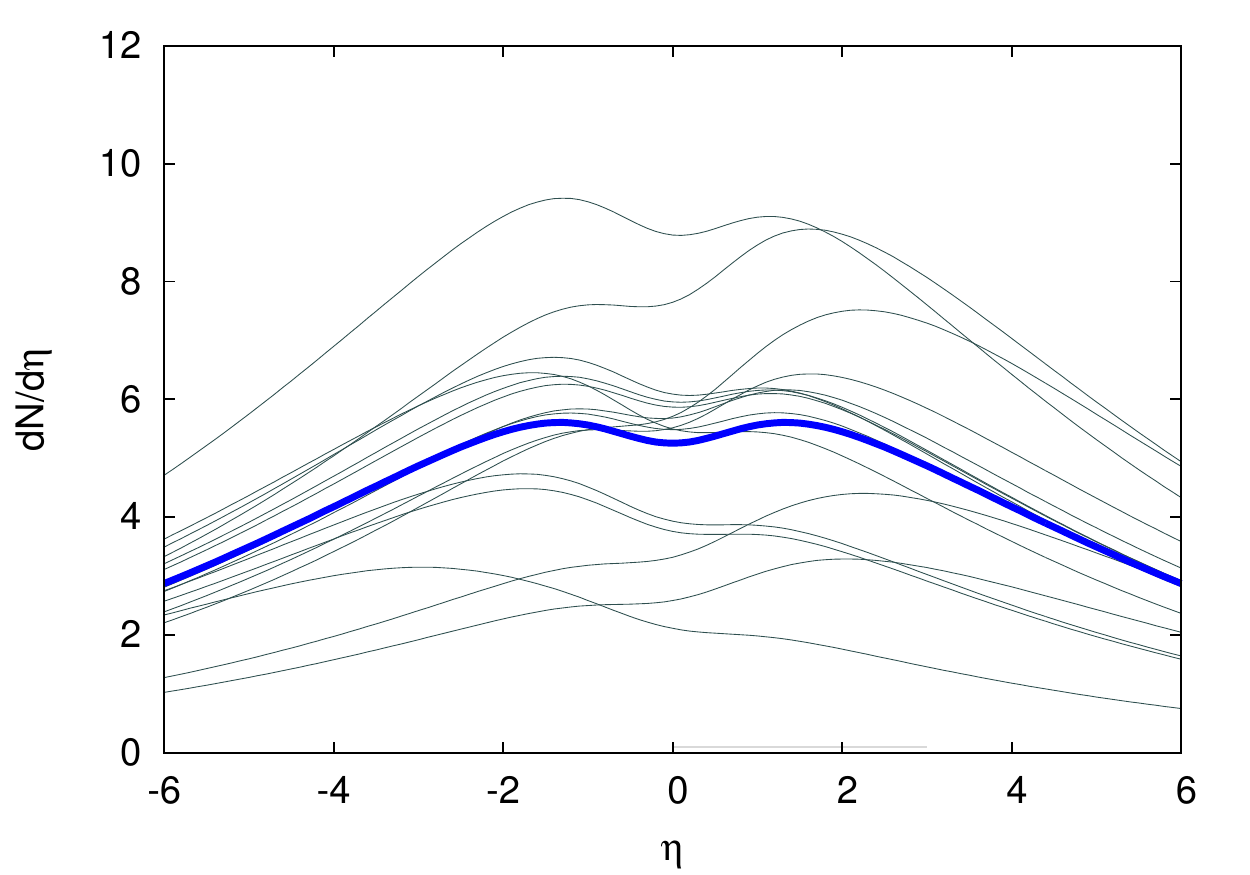}
\caption{Event-by-event charged particle rapidity distribution in p+p
collisions for $\lambda=0.32$ and $\sigma=0.5$.  
The event averaged $\left<dN/d\eta\right>$ is shown as the thicker blue curve.
The conversion between rapidity $y$ and pseudo-rapidity $\eta$ was done in the
same manner as \cite{McLerran:2015lta}.}
\label{fig:a1mc}
\end{figure}

The basic idea is that partonic fluctuations in the
proton result in a spread of saturation scales about the mean value.  Since the
saturation scale of the two protons involved in the collision fluctuate
independently one expects an asymmetric single particle rapidity distribution,
$dN/dy$, on a per-event basis. Figure~\ref{fig:a1mc} shows a model calculation
of the rapidity distribution from a subset of events drawn from the
distribution of equation~\ref{eq:Prho}.  The event averaged
$\left<dN/d\eta\right>$ is shown as the thicker blue curve.  

On a per-event basis the single particle rapidity distribution can be
characterized by fluctuations about the mean rapidity distribution
\begin{equation}
\frac{dN}{dy}=\left\langle \frac{dN}{dy}\right\rangle \left(
1 + a_0 + a_{1}y + \ldots \right)
\label{eq:dndy}
\end{equation}
In the above expression $a_0$ characterizes event-by-event fluctuations in the
multiplicity while $a_{1}$ captures the asymmetry in the rapidity distribution
\cite{Bzdak:2012tp} on a per-event basis, due in our model to the unequal
saturation scales of the colliding protons.  We should stress that strictly
speaking equation~\ref{eq:dndy} characterizes the per-event rapidity distribution in the
limit of a large number of particles when statistical fluctuations are
negligible.  In order for experiments to obtain information on the dynamical
fluctuations encoded in $a_i$ and disentangle these from statistical
fluctuations it is necessary to measure the two-particle correlation function.
By construction the event averaged $\langle a_{i}\rangle$, for $i\geq 0$ must
vanish and we therefore focus on the root-mean-square values of the
event-by-event $a_{i}$ via the  quantities $\langle a_i a_j\rangle$, which can
be extracted from the two-particle correlation function~\cite{Bzdak:2012tp}
\begin{equation}
\frac{C_{2}(y_{1},y_{2})}{\left\langle dN/dy_{1}\right\rangle \left\langle
dN/dy_{2}\right\rangle }=\left\langle a_{0}^{2}\right\rangle+\left\langle a_{0} a_{1}\right\rangle \left(y_{1}+y_{2}\right)+\left\langle a_{1}^{2}\right\rangle y_{1}y_{2} + \cdots \,,
\label{C2-simple}
\end{equation}
where 
\begin{equation}
C_{2}(y_{1},y_{2})\equiv \left\langle \frac{d^{2}N}{dy_{1}dy_{2}}%
\right\rangle -\left\langle \frac{dN}{dy_{1}}\right\rangle \left\langle 
\frac{dN}{dy_{2}}\right\rangle\,.
\label{C2-def}
\end{equation}

As is evident from figure~\ref{fig:a1mc} the per-event $a_0$ and therefore
$\langle a_0^2\rangle$ is highly sensitive to saturation scale fluctuations.
These event-by-event multiplicity fluctuations are related to the multiplicity
distribution itself, and while saturation scale fluctuations have recently been
shown necessary to explain the tail of the multiplicity distribution
\cite{McLerran:2015qxa} additional sources must also be included such as impact
parameter fluctuations, fluctuations due to conservation laws, statistical fluctuations,
etc.  Unambiguously disentangling saturation scale fluctuations from other
possible sources will require additional observables.

The asymmetric correlator $\langle a_0 a_1\rangle$ vanishes for symmetric
colliding systems, such as p+p as considered in this work.  Due to the
independently fluctuating saturation scales of the two protons an
event-by-event asymmetry with respect to $y$ is seen in figure~\ref{fig:a1mc}.
The nature of the event-by-event asymmetric rapidity distribution can be
captured by the coefficient $\langle a_{1}^{2}\rangle$ in the above expansion
of the two-particle correlation.  

One advantage of looking at rapidity asymmetric fluctuations, is that many
of the more mundane sources of fluctuations that contribute to the multiplicity
will not contribute to the coefficient $\langle a_{1}^{2}\rangle$ due to their
rapidity symmetric particle production.  This will be discussed further in
section~\ref{sec:discussion}.  

In what follows a simple model is presented, that will allow us to derive a
relation between the experimentally measured $\langle a_{1}^{2}\rangle$ and the
variance, $\sigma$, of the proton saturation scale.

\section{Rapidity distribution in proton-proton collisions}

Consider the single inclusive rapidity distribution for a generic asymmetric
nucleus-nucleus collision \cite{Kharzeev:2004if} valid outside of the
fragmentation region of the two nuclei,
\begin{equation}
\frac{dN}{dy}\propto \Sp\mathrm{Min}[Q_{1}^{2},Q_{2}^{2}]\left( 2+\ln \frac{%
\mathrm{Max}[Q_{1}^{2},Q_{2}^{2}]}{\mathrm{Min}[Q_{1}^{2},Q_{2}^{2}]}\right)\,.
\label{eq:KLN}
\end{equation}
The proportionality constant will cancel in the ratio in
equation~\ref{C2-simple} and therefore is not needed in this work.  The
rapidity dependence enters through the evolution of the saturation scale with
$y$ as
\begin{equation}
Q_{1}^{2}=Q_{o,1}^{2}e^{+\lambda y}\,,\,\,\,Q_{2}^{2}=Q_{o,2}^{2}e^{-\lambda
y}
\end{equation}%
where $Q_{o,1}$ and $Q_{o,2}$ are the \emph{initial} saturation scales in the
two projectiles before evolution to the final saturation scales $Q_{1}$ and
$Q_{2}$ respectively.  As shown in
\cite{Munier:2003vc,Munier:2003sj,Iancu:2004es,Marquet:2006xm} the dispersion
in the final saturation scales are caused by fluctuations in the low density
tail of the initial condition and can be realized by averaging over the initial
saturation momenta drawn from a Gaussian distribution in the logarithm of the
saturation scale, 
\begin{equation}
P[\rho ]=\frac{1}{\sqrt{2\pi }\sigma }\exp \left[ -\frac{\rho ^{2}}{2\sigma
^{2}}\right]\,\,\,\, {\rm where}\,\,\,\, \rho\equiv \ln\left(\frac{Q^2}{\QmeanSq{}}\right)\,.
\label{eq:Prho}
\end{equation}
Our treatment extends the work of \cite{McLerran:2015lta} to include fluctuations in {\em both} nucleons in which case the event averaging can be performed according to
\begin{equation}
\left\langle \mathcal{O}\right\rangle =\int_{-\infty }^{+\infty }d{\rho _{1}}
d{\rho _{2}}P[{\rho _{1}}]P[{\rho _{2}}]\;\mathcal{O}[{\rho _{1}},{\rho _{2}}] ,
\label{eq:O}
\end{equation}
where the two saturations scale $Q_{o,1}^2$ and $Q_{o,2}^2$, fluctuate event-by-event around the mean
saturation scales $\QmeanSq{o,1}$ and $\QmeanSq{o,2}$ respectively.  In this work we are considering symmetric proton-proton collisions and therefore $\QmeanSq{o,1}=\QmeanSq{o,2}$ which we set to be equivalent to $\QmeanSq{o}$.  By defining,
\begin{equation}
{\rho _{1}}\equiv \ln \frac{Q_{o,1}^{2}}{\QmeanSq{o}}\,,\,\,\,{\rho _{2}}%
\equiv \ln \frac{Q_{o,2}^{2}}{\QmeanSq{o}}
\end{equation}
we can re-express equation \ref{eq:KLN} as 
\begin{eqnarray}
\frac{1}{\Sp\QmeanSq{o}}\frac{dN}{dy}\propto
\begin{cases}
e^{{\rho _{1}}+\lambda y}\left( 2+{\rho _{2}}-{\rho _{1}}-2\lambda y\right) ,
& \mbox{if  }2\lambda y<{\rho _{2}}-{\rho _{1}} \\ 
e^{{\rho _{2}}-\lambda y}\left( 2+{\rho _{1}}-{\rho _{2}}+2\lambda y\right) ,
& \mbox{if  }2\lambda y\geq {\rho _{2}}-{\rho _{1}}%
\end{cases}
\label{eq:dNdy1}
\end{eqnarray}

\begin{figure}[tbp]
\includegraphics[scale=0.9]{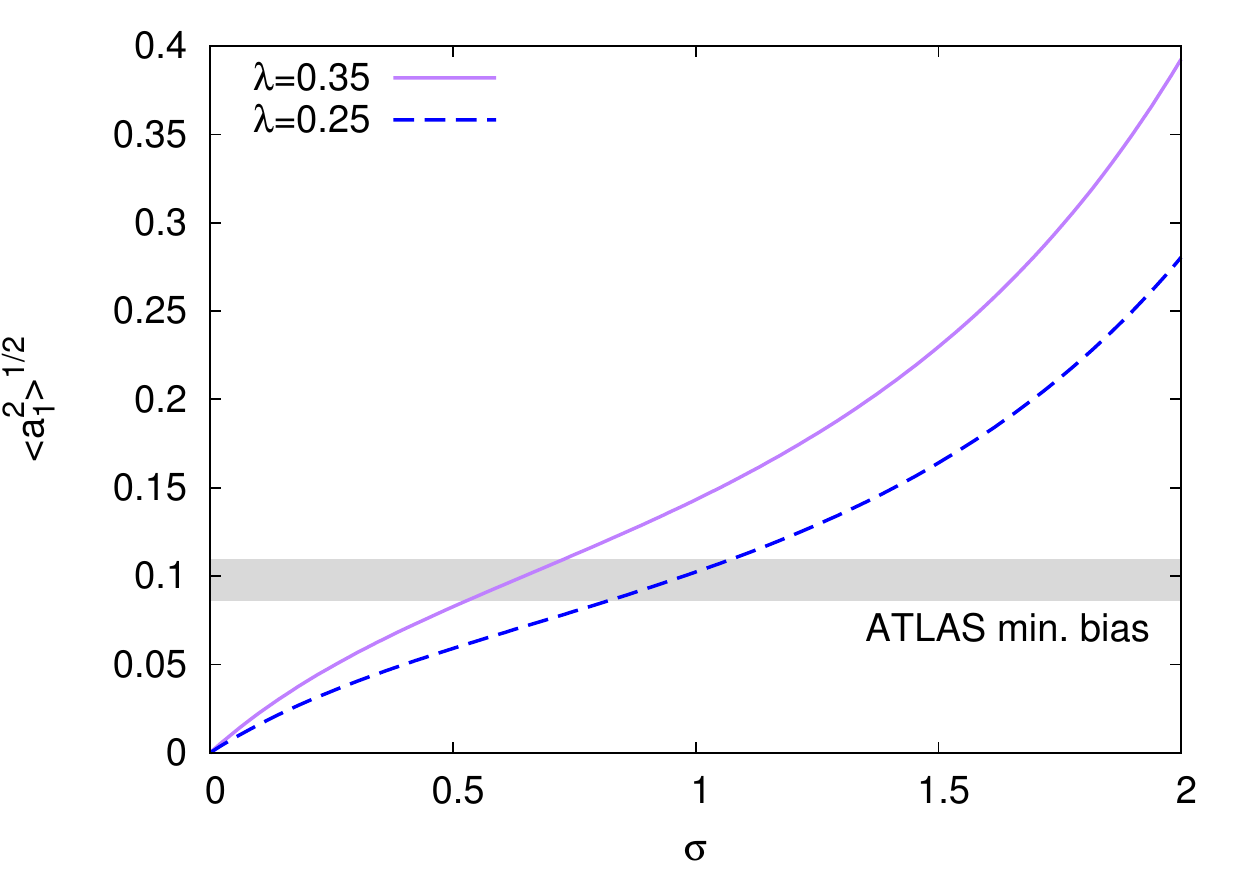}
\caption{Plot of $\sqrt{\langle a_{1}^{2}\rangle }$ from
equation~(\ref{eq:a1}) as a function of $\sigma$ for $\lambda =0.35$ 
and $0.25$. The horizontal bar is representative of the data 
on $\sqrt{\langle a_{1}^{2}\rangle}$ in minimum bias p+p collisions recently available by 
the ATLAS collaboration \cite{1395329}.  
}
\label{fig:a1data}
\end{figure}

With the above expressions in hand one can calculate the quantities $\langle dN/dy\rangle $ and $\langle d^{2}N/dy_{1}dy_{2}\rangle$.  As the final expressions are rather formidable we have included them in the Appendix.  Near mid-rapidity one can derive a rather simple expression for the coefficient $\left\langle a_{1}^{2}\right\rangle$ for minimum bias proton-proton collisions.

\begin{equation}
\left\langle a_{1}^{2}\right\rangle \simeq \dfrac{\lambda ^{2}\sigma ^{2}}{2}%
\dfrac{4\pi \left( 1+2\sigma ^{2}\right) \exp \left[ \sigma ^{2}\right] 
\mathrm{Erfc}\left[ \sigma \right] -8\sqrt{\pi }\sigma }{\left( \sqrt{\pi }%
\left( \sigma ^{2}-2\right) \mathrm{Erfc}\left[ \dfrac{\sigma }{2}\right]
-2\sigma \exp \left[ -\dfrac{\sigma ^{2}}{4}\right] \right) ^{2}} \,,
\label{eq:a1}
\end{equation}
where ${\rm Erfc}$ is the complementary error function. Equation \ref{eq:a1} is the main result of this work.  It provides a direct
relation between asymmetric rapidity fluctuations $\left\langle
a_{1}^{2}\right\rangle$ and the variance, $\sigma$, of saturation scale fluctuations in
the proton.  We plot equation~\ref{eq:a1} as a function of $\sigma$ for two representative values of $\lambda=0.25,\,0.35$.  Phenomenological fits of Deep Inelastic Scattering data at small-$x$  \cite{GolecBiernat:1998js,Praszalowicz:2012zh,Praszalowicz:2015dta} constrain $\lambda$ within this range.
The horizontal band in figure~\ref{fig:a1data} is representative of the data from minimum bias proton-proton collisions at $\sqrt{s}=13$~TeV.  The band is centered at $\left\langle
a_{1}^{2}\right\rangle=0.098$ corresponding to the ATLAS collaboration measurement~\cite{1395329}  at $\Nch=17.6$ and the thickness of the band is two standard deviations $\pm 0.012$.  We should caution the reader that the ATLAS collaboration has used very narrow centrality classes and that a more direct comparison with data would use the same centrality cuts as the experiment.  This is left to future work.

Inspection of figure~\ref{fig:a1data} shows that a value of $\sigma$ in the
range $0.5-1$ is consistent with the experimental data.  This is in qualitative
agreement with the values obtained by fitting p+p multiplicity distributions
\cite{McLerran:2015qxa} ($\sigma = 0.5$) and p+Pb rapidity  distributions
\cite{McLerran:2015lta} ($\sigma = 1.55$).  

\section{Discussion and conclusions}
\label{sec:discussion}

There are a number of caveats that should be discussed before a quantitative
extraction of $\sigma$ is undertaken. First, the model under consideration in
this work only includes intrinsic fluctuations of the proton saturation
momentum.  It is well known that in order to have quantitative agreement with
the multiplicity distribution both impact parameter and color charge
fluctuations must be included.  However, it is plausible that many sources of
fluctuations that contribute to the multiplicity have a smaller effect on the
asymmetric coefficient $\left<a_1^2\right>$.  For example, impact parameter
fluctuations from a radially symmetric proton will produce a symmetric rapidity
distribution in each event and will therefore not contribute to
$\sqrt{\left<a_1^2\right>}$.  The color charge fluctuations as included in
\cite{McLerran:2015qxa} will also not contribute -- the results of
\cite{McLerran:2015qxa} are boost invariant in each event.  More generally
color charge fluctuations, with quantum evolution, may introduce a
forward-backward asymmetry on an event-by-event basis but we expect them to be
suppressed by $1/(Q_s^2\Sp)$ where $Q_s^2$ represents the length scale of the
color charge fluctuations and $\Sp$ the collision overlap area.     

In this paper we calculated the two-particle correlation function originating
from the intrinsic fluctuations of the saturation scales in two colliding
protons. Clearly this mechanism correlates not only two particles but also
leads to multi-particle rapidity correlations. In
reference~\cite{Bzdak:2015dja} it was argued that higher order correlation
functions naturally remove unwanted short-range rapidity correlations, e.g.,
resonance decays.  In this work we focused on correlations after the
experimental subtraction of the short range component.

\bigskip

In conclusion, we demonstrated that intrinsic fluctuations of the proton
momentum saturation scale leads to an event-by-event asymmetric single particle
distribution, $dN/dy$, and consequently to nontrivial rapidity correlations. We
extracted the asymmetry coefficient $\sqrt{\left<a_1^2\right>}$ that directly
measures the width of saturation scale fluctuations, that is,
$\sqrt{\left<a_1^2\right>}$ is roughly proportional to $\lambda\sigma$, the
width of Gaussian fluctuations in logarithm of the saturation scale, see
figure~\ref{fig:a1data}. When compared to the preliminary ATLAS data we
conclude that $\sigma$ is of the order of $0.5-1$.

\bigskip

\vspace{\baselineskip} 
\noindent \textbf{Acknowledgments:} \newline
{}\newline
We thank J.~Jia and M.~Praszalowicz for valuable correspondence and L.~McLerran for his encouragement. AB was supported by the Ministry of Science and Higher Education (MNiSW), by founding from the Foundation for Polish Science, and by the National Science Centre, Grant No. DEC-2014/15/B/ST2/00175, and in part by DEC-2013/09/B/ST2/00497.

%

\appendix

\section{Equations for the one- and two-particle rapidity distributions}

In this section we collect analytic expressions for the one- and two-particle rapidity distributions and discuss the derivation of equation~\ref{eq:a1}.  Using equations~\ref{eq:Prho}, \ref{eq:O}, \ref{eq:dNdy1} the event averaged multiplicity distribution is
\begin{align}
\frac{1}{\Sp\QmeanSq{o}}\left\langle \frac{dN}{dy}\right\rangle & =%
\frac{\sigma }{\sqrt{\pi }}\exp \left[ \frac{\sigma ^{2}}{4}-\frac{\lambda
^{2}y^{2}}{\sigma ^{2}}\right] +\left( 1+\lambda y-\frac{\sigma ^{2}}{2}%
\right) \exp \left[ \frac{\sigma ^{2}}{2}-\lambda y\right] \mathrm{Erfc}%
\left[ \frac{\sigma }{2}-\frac{\lambda y}{\sigma }\right]   \notag \\
& +\left\{ y\rightarrow -y\right\} 
\end{align}
where ${\rm Erfc}=1-{\rm Erf}$ is the complementary error function.  The two particle rapidity correlation is more cumbersome and the final expression reads
\begin{eqnarray}
\frac{1}{\left(\Sp\QmeanSq{o}\right)^2}\left\langle \frac{d^{2}N}{dy_{1}dy_{2}} \right\rangle =
\begin{cases}
N_{2}(y_{1},y_{2}), & \mbox{if }y_{2} > y_{1} \\ 
N_{2}(y_{2},y_{1}), & \mbox{if }y_{2}\leq y_{1}
\end{cases}
\end{eqnarray}
where we defined
\begin{align}
& N_{2}(y_{1},y_{2})=2\left( \frac{\sigma ^{2}}{2}+(\lambda y_{2}+1)(\lambda
y_{1}-1)\right) \mathrm{Erf}\left[ \frac{\lambda y_{1}}{\sigma }\right] \exp %
\left[ \sigma ^{2}+\lambda (y_{1}-y_{2})\right]   \notag \\
& +2\left( \sigma ^{4}-\sigma ^{2}\left( \lambda (y_{1}+y_{2})+\frac{3}{2}%
\right) +(\lambda y_{2}+1)(\lambda y_{1}+1)\right) \mathrm{Erfc}\left[
\sigma -\frac{\lambda y_{1}}{\sigma }\right] \exp \left[ 2\sigma
^{2}-\lambda (y_{1}+y_{2})\right]   \notag \\
& -\frac{4\sigma }{\sqrt{\pi }}\left( \frac{\sigma ^{2}}{2}+\lambda
y_{1}-1\right) \exp \left[ \sigma ^{2}+\lambda (y_{1}-y_{2})-\frac{\lambda
^{2}y_{2}^{2}}{\sigma ^{2}}\right]   \notag \\
& +\left\{ (y_{1},y_{2})\rightarrow (-y_{2},-y_{1})\right\} %
\phantom{\left(\frac{sigma^2}{2}\right)}
\end{align}

Equation~\ref{eq:a1} is an approximation strictly valid in the limit $\lambda Y \rightarrow 0$.  
Most generally $C_{2}(y_{1},y_{2})$ can be expressed in terms of orthogonal polynomials \cite{Bzdak:2012tp}
\begin{equation}
\frac{C_{2}(y_{1},y_{2})}{\left\langle dN/dy_{1}\right\rangle \left\langle
dN/dy_{2}\right\rangle }=\sum_{i,k=0}\left\langle
a_{i}a_{k}\right\rangle T_{i}(y_{1})T_{k}(y_{2}),  
\label{eq:C2-full}
\end{equation}%
where following the notation employed in \cite{Jia:2015jga,1395329}
\begin{equation}
T_{k}(y)=Y\sqrt{\frac{2k+1}{3}}P_{k}(y/Y)\,,\,\,\,\,\, P_{k}(x)=\frac{1}{2^{k}k!}\frac{d^{k}}{dx^{k}}(x^{2}-1)^{k}\,.
\end{equation}%
In the above expression $P_k(x)$ are Legendre polynomials ({\em e.g.}
$P_0(x)=1\,,P_1(x)=x\,,\ldots$ ) and the rapidity distribution is constrained
to be measured in the interval $-Y \leq y \leq +Y$. 
Instead of using the full expression
\begin{equation}
\left\langle a_{1}^{2}\right\rangle =\left( \frac{3}{2Y^{3}}\right)
^{2}\int_{-Y}^{Y}dy_{1}dy_{2}\frac{C_{2}(y_{1},y_{2})}{\left\langle
dN/dy_{1}\right\rangle \left\langle dN/dy_{2}\right\rangle }%
T_{1}(y_{1})T_{1}(y_{2}),
\end{equation}
we have used a Taylor series expansion around mid-rapidity (see Eq. \ref{C2-simple})
\begin{equation}
\left\langle a_{1}^{2}\right\rangle \simeq \left.\frac{d}{dy_{1}}\frac{d}{dy_{2}}%
\frac{C_{2}(y_{1},y_{2})}{\left\langle dN/dy_{1}\right\rangle \left\langle
dN/dy_{2}\right\rangle }\right|_{y_{1}=y_{2}=0}.
\end{equation}
This results in $\left\langle a_{1}^{2}\right\rangle -\sqrt{21}\left\langle
a_{1}a_{3}\right\rangle +(21/4)\left\langle a_{3}^{2}\right\rangle +...$,
however, the higher components 
are small for small values of $\lambda Y$.  We have checked and even for the acceptance of $-2.4\leq y \leq 2.4$ these corrections amount to less than 10\% on equation~\ref{eq:a1}.

\end{document}